# The Calculation of Kinetic and Static Friction Coefficient and Friction Graph Analysis Using Arduino

**Atakan Çoban[1] and Seher Boyacı[2]**

[1]Yeditepe University, Faculty of Arts and Sciences, Department of Physics, Ataşehir, Istanbul, Turkey.
[2]Physics Education Department, Dokuz Eylül University, Education Faculty of Buca, İzmir, Turkey.

E-mail: atakancoban39@gmail.com

**Abstract**

In this study, with the help of the Arduino UNO and Load-cell force sensor, a simple experimental material has been developed to calculate kinetic and static friction coefficients and analyse the friction force in detail. The system with a force sensor mounted on it is placed on the plane. With the help of the rope attached to the force sensor, a force, whose magnitude increased over time, was applied on the sensor. After the force applied was large enough to move the system, the force was applied to move the system at a constant speed. Since the net force acting on the system is zero while the system is stationary and moving with a constant speed, in these cases the magnitudes of force read through the sensor are equal to the static and kinetic friction forces acting on the object. Using these static and kinetic friction forces values, the maximum value of the static friction coefficient and the kinetic friction coefficient were calculated as 0.484 and 0.313, respectively.

Keywords: Physics education, Arduino, Kinetic friction, Static friction, STEM.

### 1. Introduction

One of the main purposes of educational activities today is to provide students with 21st century skills[1]. At this point, STEM education has begun to be researched and used frequently. STEM education aims to teach Science, Technology, Engineering and Mathematics disciplines together in an interdisciplinary education environment[2]. Applications for technology and engineering disciplines in STEM training are important parts of the process. Circuit setup and programming activities can be done using Arduino in this technology and engineering applications[3]. There are many studies where Arduino is used in physics applications[4-6].

### 2. Theory

The subject of friction force is included in physics courses for many age groups[7,8]. The friction force is equal to the multiplication of the friction coefficient (μ) of the surface of the body and the normal force (N) acting on the body. If the applied force is not large enough to set the object in motion, the friction force is called static friction force and its magnitude is equal to the magnitude of the applied force. Static friction force can take values between 0 and $f_{smax}$. The maximum value that the static friction force can take is calculated with the equation $f_{smax}=\mu_s N$. $\mu_s$ is the maximum value of coefficient of static friction and can be calculated with the equation

$$\mu_s = \frac{f_{smax}}{N} \quad (1)$$

If the applied force moves the object, the friction force acting on the object takes a fixed value, which is called the kinetic friction force. $f_k=\mu_k n$ is the equation for calculating the magnitude of the kinetic friction force. $\mu_k$ is the kinetic friction coefficient and can be calculated with the equation

$$\mu_k = \frac{f_k}{N} \qquad (2)$$

According to the law of inertia, if the object has a constant speed, the net force acting on it is zero. In the stationary state of the object, the force applied on the object is equal to the static friction force, and when it moves at a constant speed, the applied force is equal to the kinetic friction force.

### 3. Experimental Setup

In this study, a Load-cell force sensor, a HX711 amplifier, an Arduino UNO, wooden surface, connection cables, and a computer were used. Loadcell sensor is used to measure force with different sensitivities[9]. Since the signals coming over the sensor are small, they are amplified with the help of HX711 and sent to the Arduino. The connection of the sensor to HX711 and HX711 to the Arduino is shown in Figure 1.

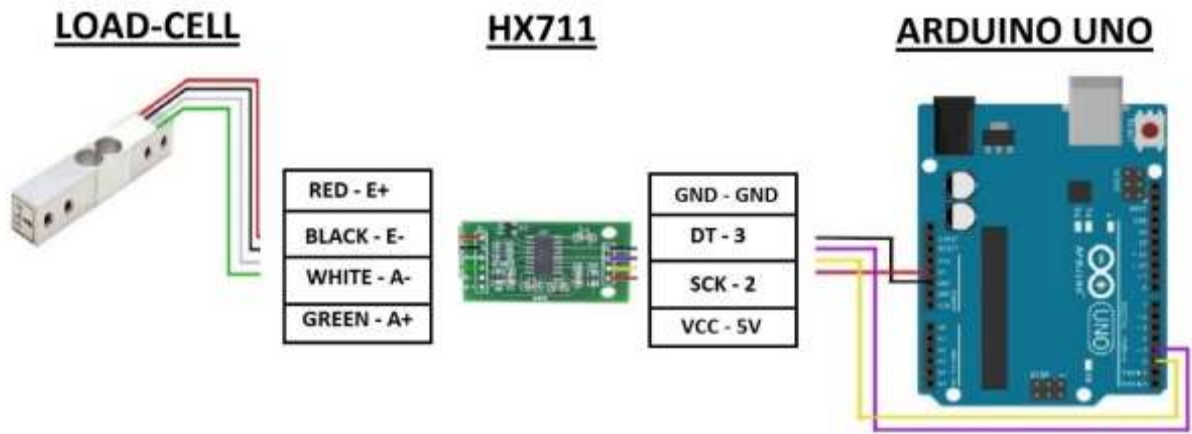

**Figure 1.** Connecting between the Load-cell, the HX711 and the Arduino UNO

The force sensor is mounted vertically on the board and the necessary connections have been made with Arduino as in Figure 2.

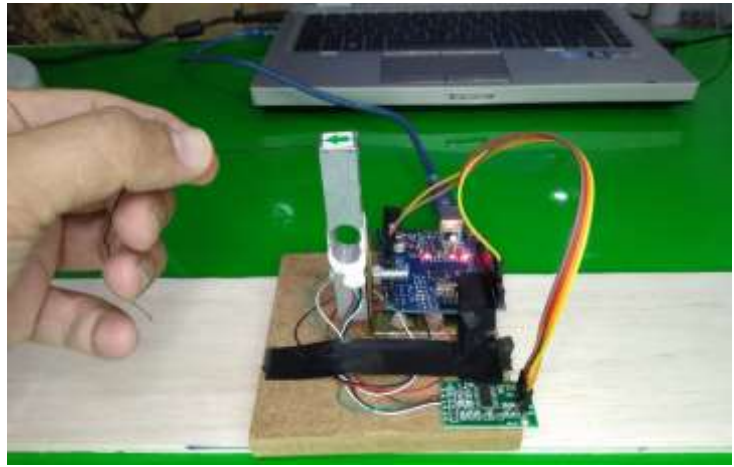

**Figure 2.** Experimental setup.

During the data collection process, force was applied on the rope attached to the force sensor as in Figure 2. The applied force was gradually increased, starting at zero. The system started to move after a certain value of the force. From this moment on, the force has been applied to move the system at an almost constant speed. It was determined with the help of observation that the system was moving at a constant speed. The force is applied in such a way that the system has approximately equal displacements in an equal time interval. In the part where the system moves, it should be ensured that the cable has freedom of movement in order to prevent the effect of the cable connected to the Arduino computer on the movement.

## 4. Data Analysis

While the force acts on the sensor system, the net force is equal to the difference between the tensile force T and the friction force f. When the system is inert, that is, when stationary or moving at a constant velocity, the net force $F_{net}$ on it is zero. Therefore, in these cases, the force stretching the rope is equal to the friction force. When the system is stationary and moving at a constant speed, the net force acting on it will be zero, so the values read by the force sensor in these processes are equal to the static and kinetic friction forces, respectively. Force and time data were taken during this process. The screenshot of the data read on the Arduino IDE program is shown in Figure 3.

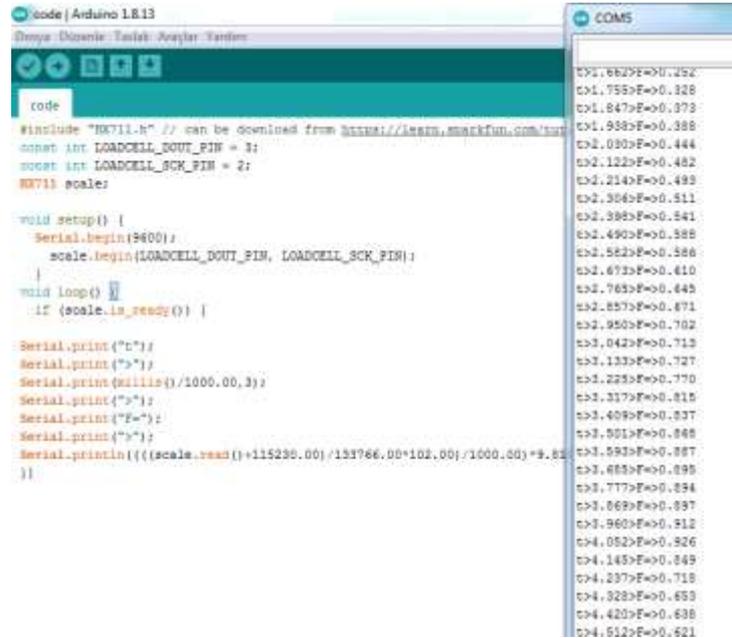

**Figure 3.** Screenshot of the received data. The values on the left are time values in seconds, and the values on the right are tensile force values in Newton.

It can be seen that the frictional force value increases with time in accordance with the theory, and after a certain maximum value it decreases a little and gets a constant value. When plotting the data, an identical graph with the theoretical graph is obtained as in figure 4.

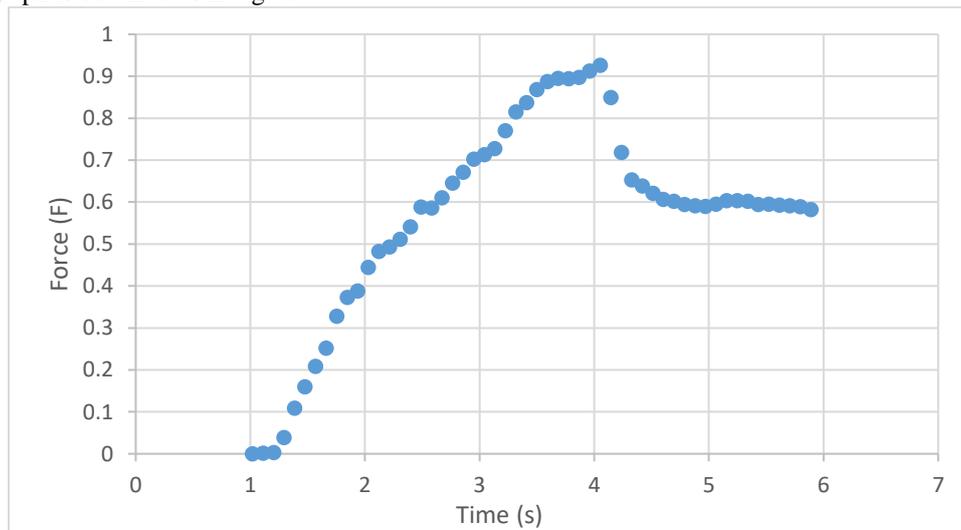

**Figure 4.** Graph of the change of force acting on the object with time.

Since the force measured by the sensor during the data collection process is equal to the friction force, the graph shown in Figure 4 is identical to the time-dependent graph of the friction force. Here, it is seen that a graph suitable for the theoretical graph[11]. It is seen that the maximum value of the static friction force is 0.926 N and the average kinetic friction force is 0.599 N. The total weight of the force sensor and the wooden block, which is also equal to its normal force, was found as $N = 1.913\ N$. Static friction coefficient is calculated as $u_s = 0.484$ and kinetic friction coefficient is calculated as $u_k = 0.313$ by using 1 and 2 equations. In order to test the accuracy of the obtained results, a widely used method was used to calculate static and kinetic friction coefficients. The slope of the surface was increased slowly and the angle value when the object started to move was determined as $24^0$. The tangent of this angle value, 0.445, is directly equal to the static friction coefficient. While increasing the slope of the surface, a small thrust force was applied on the object at every inclination value and the angle value at the moment the object started to move with a constant speed was determined as $19^0$. The tangent of this angle value, 0.344, is also equal to the kinetic friction coefficient. These results and the results obtained in the study are very close.

### 5. Conclusion

The total cost of the basic materials used in the experiment was approximately $6. Considering the access of almost everyone to computers today, the results that can be obtained from very high-cost experimental sets have been achieved instead with tools obtained at very affordable prices in this study. In this study, there are gains from technology, engineering, mathematics and science, and this is an example of the STEM education approach. In addition, thanks to the method followed in the study, it is possible to calculate both static friction force and kinetic friction force simultaneously on the same system and plot them. This application, where the difference between static and kinetic friction can be clearly seen with such an easy application, will both facilitate the learning of students and increase their interest in the lesson.